\documentclass[lettersize,journal]{IEEEtran}
\usepackage{amsmath}
\usepackage{amssymb}
\usepackage{amsfonts}

\DeclareMathOperator*{\argmin}{arg\,min}
\DeclareMathOperator{\Softmax}{Softmax}
\usepackage{algorithm}
\usepackage{algpseudocode}
\algnewcommand{\Initialize}[1]{%
  \State \textbf{Initialize:}
  \Statex \hspace*{\algorithmicindent}\parbox[t]{.8\linewidth}{\raggedright #1}
}
\usepackage{array}
\usepackage[caption=false,font=normalsize,labelfont=sf,textfont=sf]{subfig}
\usepackage{textcomp}
\usepackage{stfloats}
\usepackage{url}
\usepackage{verbatim}
\usepackage{graphicx}
\usepackage{cite}
\usepackage{fancybox, graphicx}
\usepackage{tabularx}
\usepackage[export]{adjustbox}
\usepackage{makecell}
\usepackage{array}
\usepackage{balance}
\usepackage{hyperref}
\usepackage{color}
\usepackage{multirow}
\usepackage{booktabs} 
\usepackage{subfig}

\usepackage{orcidlink}

\begin{document}

\title{Temporal Graph Memory Networks For Knowledge Tracing}

\author{\IEEEauthorblockN{Seif Gad\IEEEauthorrefmark{1},
Student Member, IEEE,
Sherif Abdelfattah\IEEEauthorrefmark{1}\IEEEauthorrefmark{2} and
Ghodai Abdelrahman, Member, IEEE}\\
}
\maketitle
\def\thefootnote{*}\footnotetext{Denotes equal contribution.}\def\thefootnote{\arabic{footnote}}
\def\thefootnote{\IEEEauthorrefmark{2}}\footnotetext{Corresponding author: dr.shixo@gmail.com}\def\thefootnote{\arabic{footnote}}
\markboth{Journal of \LaTeX\ Class Files,~Vol.~14, No.~8, August~2021}%
{Shell \MakeLowercase{\textit{et al.}}: A Sample Article Using IEEEtran.cls for IEEE Journals}


\maketitle

\begin{abstract}
Tracing a student's knowledge growth given the past exercise answering is a vital objective in automatic tutoring systems to customize the learning experience. Yet, achieving this objective is a non-trivial task as it involves modeling the knowledge state across multiple knowledge components (KCs) while considering their temporal and relational dynamics during the learning process. Knowledge tracing methods have tackled this task by either modeling KCs' temporal dynamics using recurrent models or relational dynamics across KCs and questions using graph models. Albeit, there is a lack of methods that could learn joint embedding between relational and temporal dynamics of the task. Moreover, many methods that count for the impact of a student's forgetting behavior during the learning process use hand-crafted features, limiting their generalization on different scenarios. In this paper, we propose a novel method that jointly models the relational and temporal dynamics of the knowledge state using a deep temporal graph memory network. In addition, we propose a generic technique for representing a student's forgetting behavior using temporal decay constraints on the graph memory module. We demonstrate the effectiveness of our proposed method using multiple knowledge tracing benchmarks while comparing it to state-of-the-art methods.  
\end{abstract}


\section{Introduction}
\label{sec:introduction}
Teaching has always been a vital skill for the human race to transfer knowledge from one generation to another. During a teaching procedure, a human teacher would utilize their domain knowledge about the subject and its underlying learning concepts to customize the learning experience for individual students based on their knowledge progress. Such a customization process demands a continuous tracing of a student's knowledge growth using probing means such as asking questions. Yet this begs the question: \emph{Could a computational model trace a student's knowledge growth given their exercise answering history?} The knowledge tracing (KT) problem aims at answering such a question~\cite{corbett1994knowledge, AbdelrahmanWNSurvey23} to accurately predict the probability of correctly answering a recent question given the past exercise answering history and information on relationships between questions and knowledge components (KCs) (i.e., learning concepts) in a subject. 

\begin{figure}[t]
\centerline{\includegraphics[width=9.5cm, height=6cm]{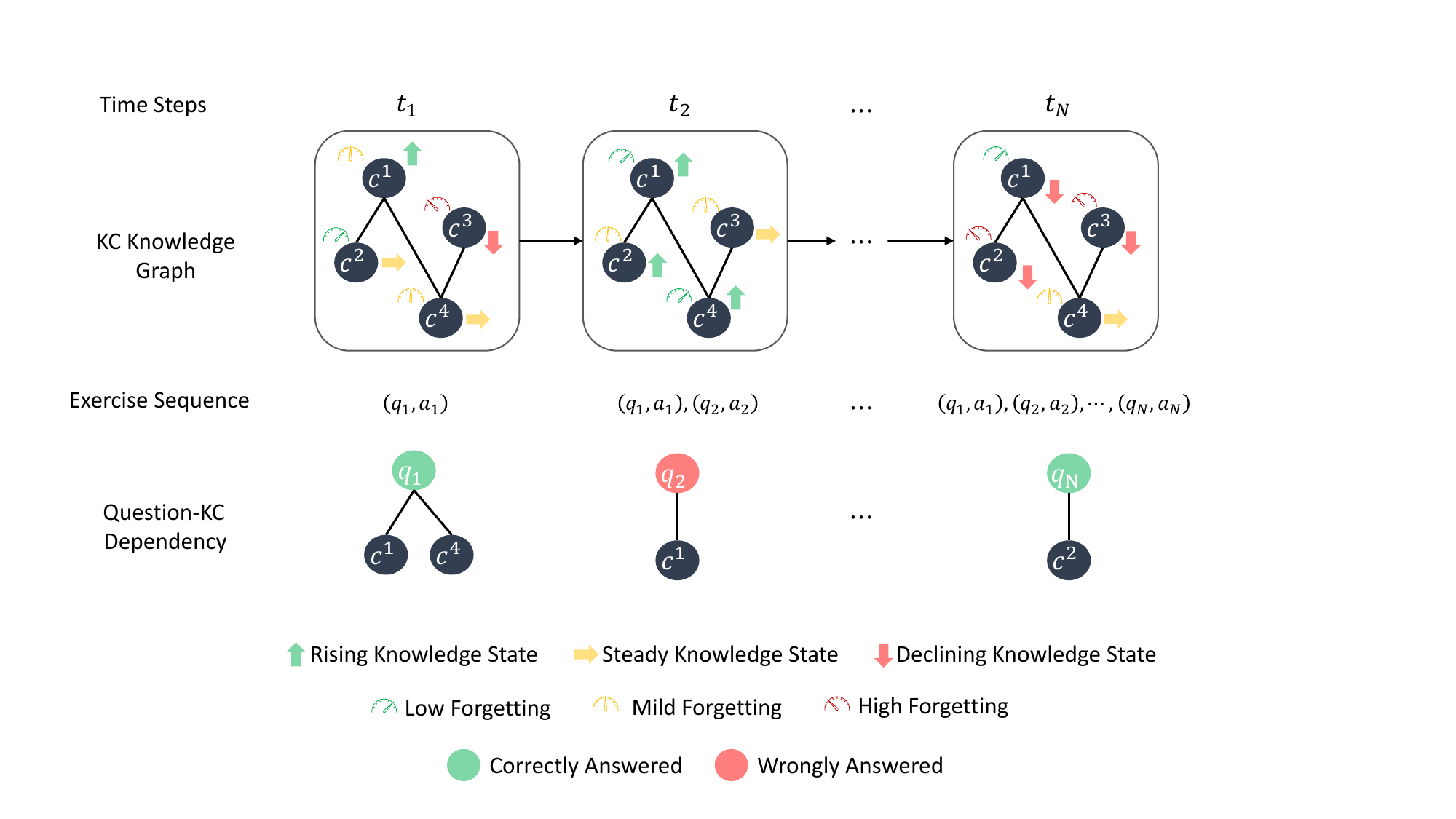}}
\caption{An example for the temporal and structural dynamics involved in the Knowledge Tracing (KT) problem.}
\label{fig:money_fig}
\end{figure}

To further clarify the semantics of a KT context and the complexity of their dynamics, we provide an illustrative example in Figure~\ref{fig:money_fig}. At the top level, we have time steps representing the temporal nature of the KT problem. The second level depicts a student's latent knowledge state as a KC-KC relationship graph with four unique KCs in the example task. We note that a knowledge state for each KC could temporarily fluctuate over multiple values, including rising, steady, or declining. Another factor that impacts the knowledge state for each KC is the student's forgetting behavior, which is described as a temporal memory decay effect~\cite{forget_NagataniZSCCO19,forgetting_15,carpenter2008effects,bjork2014forgetting} usually governed by a KC practice frequency. Forgetting behavior, represented by gauges in the second level, and relationships between KCs and questions govern the knowledge state dynamics. For example, correctly answering $q_1$ directly impacts knowledge and forgetting states of concepts ($c^1,c^4$) and indirectly impacts their related concepts ($c^2,c^3$). Addressing the KT problem involves multiple challenges, including 1) modeling knowledge state dynamics across multiple KCs, 2) counting for a student's forgetting behavior impact on predicting the knowledge state, and 3) effectively capturing the relationships between questions and KCs during the knowledge state modeling.

There have been many attempts to tackle the KT problem from a computational modeling perspective. Statistical approaches~\cite{corbett1994knowledge,Baker2008,Pardos_2011,Yudelson_2013,gunzelmann2004knowledge} followed Bayesian estimation techniques to quantify the probability of correctly answering a given question. However, they usually tend to oversimplify the problem representation (e.g., assuming only one KC in the task) for a tractable posterior computation. Alternatively, deep learning (DL) approaches~\cite{DKT2015,DKVMN17} benefit from recent advances in deep neural networks~\cite{lecun2015deep,Deng14_DL,DGL_Survey20} to embrace more realistic assumptions (e.g., multiple KCs and KCs-questions dependencies) of the KT problem. The DL approaches could be generally decomposed into two broad categories depending on the type of deep models utilized, including deep sequence models~\cite{DKT2015,DKVMN17,SKVMN,AKT20,atDKT23} and deep graph models~\cite{GKT19,TongLHHCLM020}. Despite being effective, the previous DL approaches tackle the KT problem by either considering the sequence context from the exercise answering sequences or the KC-question relationship context from dependency graph representation. Few approaches~\cite{GIKT20,abdelrahman2022deep} considered both sequence and graph contexts during the answer prediction. Yet, they modeled separate representations for sequence and graph contexts and used handcrafted features~\cite{abdelrahman2022deep} to represent a student's forgetting behavior. 

This paper proposes a novel model called \emph{Temporal Graph Memory network} (TGMN) that combines sequence and graph contexts into a joint representation using a novel graph memory structure. TGMN follows a key-value memory graph design to capture relationships between KCs using soft edges inferred from similarities in their key embeddings while tracking the knowledge growth for each KC using value embeddings. In addition, we propose a novel way to count for a student's forgetting behavior without depending on handcrafted features by using a temporal memory decay technique that is inspired by recent advances in deep sequence models to deal with long-temporal contexts~\cite{NotAllMemSame21}. In summary, our contribution points could be summarised as follows:

\begin{itemize}
    \item We propose a novel temporal graph memory network (TGMN) that could capture both exercise answering temporal context and structural relationships among KCs in a joint representation.
    \item We propose an extension of the graph convolution operation to distill information from the temporal graph memory module.
    \item We propose a generic way to count for a student's forgetting behavior during the answer prediction without needing handcrafted features using a temporal memory decay technique.
    \item We presented a comprehensive experimental analysis comparing our proposed model to the state-of-the-art KT models across multiple KT datasets. Also, we analyze the impact of different building blocks on the model's performance using an ablation study. 
\end{itemize}

The rest of this paper is organized as follows. Section~\ref{sec:pd}
introduces the knowledge tracing problem definition. Section~\ref{sec:method} elaborates the details of the proposed
Temporal Graph Memory Network (TGMN) model for knowledge tracing. Section~\ref{sec:exp} depicts the experimental setup, evaluation results, and performance analysis. Section~\ref{sec:implementation} illustrates the TGMN implementation details. Section~\ref{sec:rl} reviews the related work for tackling the knowledge tracing problem. Finally, Section~\ref{sec:conclusion} concludes the study and highlights future work directions.

\section{The Knowledge Tracing Problem}
\label{sec:pd}
The knowledge tracing problem could be formulated as given a set of questions $Q=\{q^1,q^2,\dots,q^L\}$, a set of knowledge components (KCs) $C=\{c^1,c^2,\dots,c^N\}$, the relationships between questions and KCs as a bipartite graph $G=(Q,C,E)$ where $E$ denotes the edges of the graph, and a student's exercise practice history represented as sequence $X=(\{q_1,a_1\},\{q_2,a_2\},\dots,\{q_{t-1},a_{t-1}\})$, where $q_j\in{Q}$ and $a_j\in\{0,1\}$, we seek to predict the probability of correctly answering the question at the current time step $q_t$ defined as $p(a_t=1|q_t,X,G)$.

We assume that the ability of a student to answer a given question correctly is governed by a latent mastery state $m_t\in\mathbb{R}^{d_m}$, where $d_m$ is the dimension of the state. The mastery state $m_t$ could be estimated via a parametric model given information in a KT context $E_\omega:q_t,X,G\to{m_t}$, where $\omega$ is the parameter set of the state estimator $E$. Assuming that the state $m_t$ encapsulates all the relevant information from the graph $G$ and the exercise sequence $X$, we re-formulate the prediction objective to be conditioned on the question tag $q_t$ and the state $m_t$ as $p(a_t=1|q_t,m_t)$. Our optimization objective to tackle the KT problem is as follows:

\begin{equation}
    \argmin_{\theta\in\Theta,\omega\in\Omega}\quad\mathcal{L}(a_t,F_\theta(q_t,E_\omega(q_t,X,G))
\end{equation}

\noindent where $\mathcal{L}$ is a prediction loss function, $F_\theta:q_t,m_t\to\hat{a}_t,\,\hat{a}_t\in\{0,1\}$ is an answer prediction model parametrized with $\theta$, $\Theta$ is the parameter search space for $F_\theta$, and $\Omega$ is the parameter search space for $E_\omega$.

\section{proposed Method}
\label{sec:method}
\begin{figure*}[htbp]
\centerline{\includegraphics[scale=0.55]{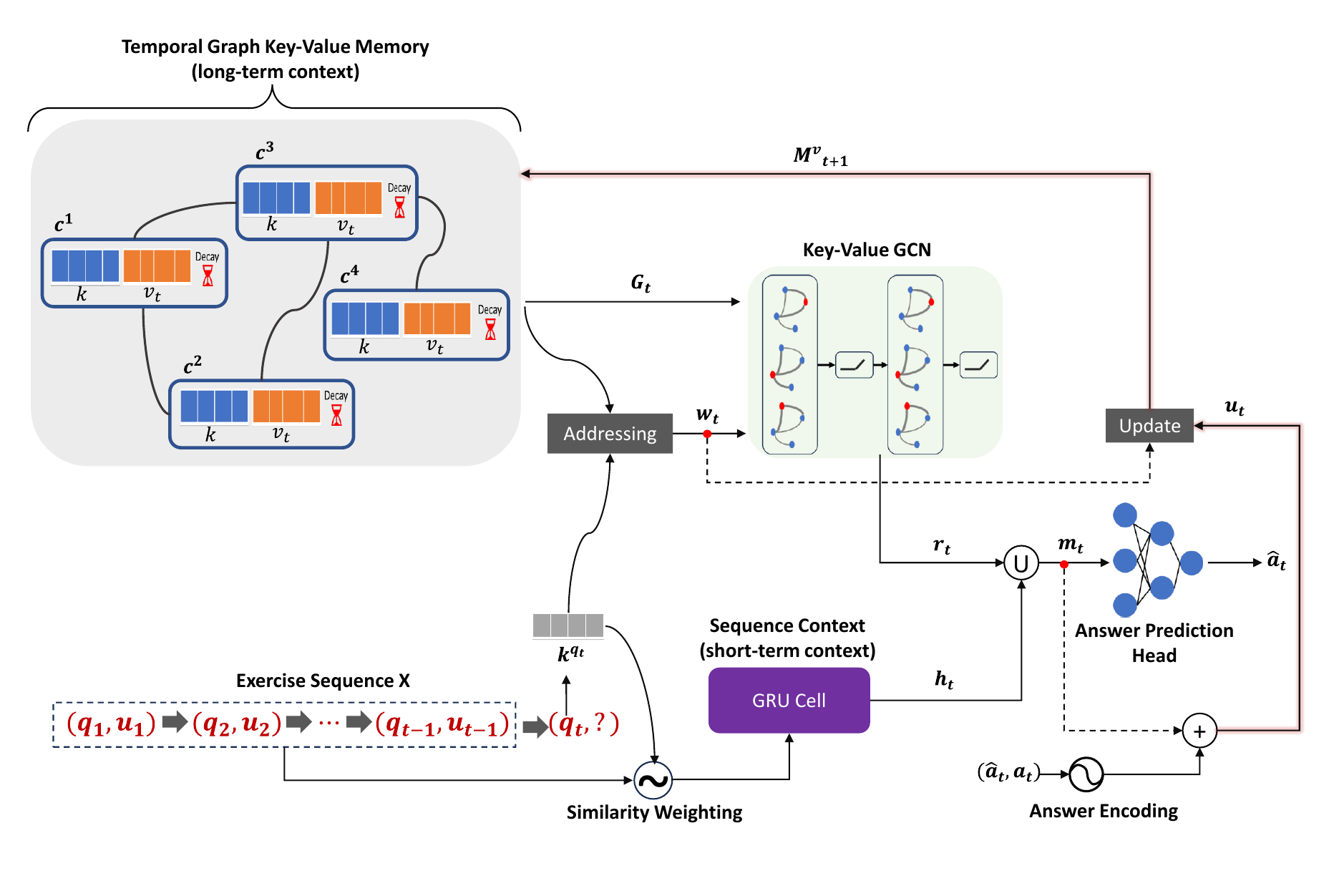}}
\caption{Architecture of the temporal graph neural network (TGMN) model for knowledge tracing. Knowledge state estimation is performed by fusing long-term knowledge context $r_t$ (across all practice history) by addressing the temporal graph memory and the short-term knowledge context $h_t$ distilled from the current exercise sequence using a GRU cell. Afterward, the estimated knowledge state $m_t$ is fed into an answer prediction head to predict the correct answer probability for question $q_t$. We update the temporal graph memory (highlighted red arrows) guided by the status of the prediction error.
}
\label{fig:main_framework}
\end{figure*}

In this section, we introduce the details of our temporal graph memory network (TGMN) model for knowledge tracing. Specifically, five main building blocks are involved: the temporal graph key-value memory, a key-value graph convolution network (GCN), the sequence context memory cell, the answer prediction head, and the memory update module. Figure~\ref{fig:main_framework} depicts a block diagram for the proposed model. The temporal graph memory stores a student's knowledge state over existing KCs in the task across multiple answering sessions, forming a long-term knowledge context where each node represents key-value embeddings for a KC. A key embedding is static and reflects the identity of a KC; we show later in Section~\ref{sec:embedding_learning} an effective way to learn such key embeddings following a self-supervised learning objective, whereas the value embedding is dynamic and represents the knowledge mastery level for a KC. We learn value embeddings in an end-to-end manner with the answer prediction task. 

Given a new question $q_t$, we address the temporal graph memory by attending the question's key embedding with KC key embeddings to generate a KC relevancy vector $w_t$. Similar to the KC key embeddings, we show in Section~\ref{sec:embedding_learning} our way of learning question key embeddings in a self-supervised learning manner. The key-value graph convolution network extracts the relevant knowledge state aspects to $q_t$ into a read vector $r_t$ by leveraging spatiotemporal features from the temporal graph memory while attending to the relevancy vector $w_t$. We count for short-term dynamics in the knowledge state within the current answering session by feeding the mastery states of past questions weighted by their relevancy to $q_t$ into the sequence context memory cell to output the sequence context vector $h_t$. Afterward, we concatenate the read vector $r_t$ representing the long-term context with the short-term context $h_t$ to form the combined mastery state $m_t$ of $q_t$. We input $m_t$ to a feed-forward prediction head to predict the correct answer probability $\hat{a_t}$. Finally, we update the temporal graph memory given the status of the answer prediction $(\hat{a_t},a_t)$, the estimated mastery state $m_t$, and the KC relevancy vector $w_t$ of question $q_t$. To count for the impact of a student forgetting, we introduce a novel temporal decay technique within the temporal graph memory update module that considers the previous practice attempts and their relevancy to a given KC. We introduce the details for each of the five building blocks in the remainder of this Section.

\subsection{Learning Question-KC Embedding Space}
\label{sec:embedding_learning}
Many KT datasets provide relationships between questions and KCs in a form that a bipartite graph could represent. We exploit this given information through a self-supervised objective to learn effective representations for questions and KCs. Given a bipartite KC-question graph, we target predicting the distance between two nodes from the same type (e.g., questions) represented by the number of hops over the other type (e.g., KCs). The assumption behind this objective is that a good node representation shall effectively capture similarities between nodes quantified by the number of hops separating them apart in the graph.

\begin{figure}[htbp]
\centerline{\includegraphics[width=9.5cm, height=3cm]{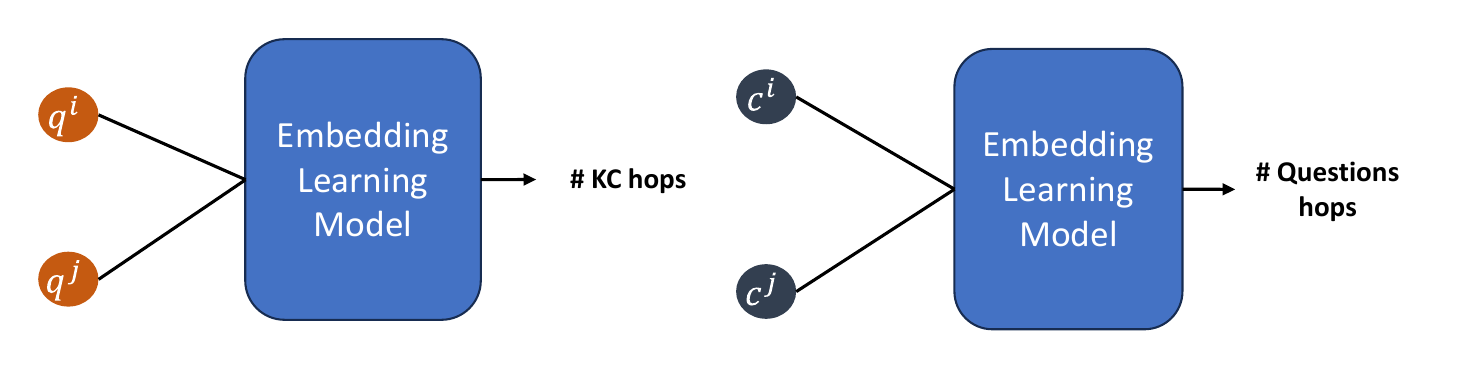}}
\caption{Graph-based pretraining tasks used to learn useful questions and KCs embeddings. The left figure presents the objective of predicting the number of KC hops between a pair of question nodes. The right figure shows the objective of predicting the number of question hops given a pair of KCs. }
\label{fig:pretraining}
\end{figure}

As shown in Figure~\ref{fig:pretraining}, the input to our embedding learning model is either one-hot encodings for two sample nodes from the same type in the case of KT datasets that do not share text tags for KCs and questions or the BERT~\cite{bert_paper,reimers2019sentence} text embeddings in the case of datasets that share it. The ground truth on the minimum number of hops between two nodes is extracted from the question-KC bipartite graph using the \emph{Dijkstra algorithm}~\cite{algo_intro_2009} following the \emph{NetworkX} Python package's~\footnote{\href{https://networkx.org/documentation/stable/reference/algorithms/generated/networkx.algorithms.shortest_paths.weighted.dijkstra_path.html}{NetworkX Dijkstra implementation}} implementation. We show the effectiveness of this node representation learning method in Section~\ref{sec:exp}.

\subsection{Temporal Graph Key-Value Memory}
The temporal graph key-value memory module aims to capture the spatiotemporal dynamics between existent KCs through key-value representations. For the rest of this paper, we call that memory module TGM, while we refer to our whole model, including all the five building blocks, as TGMN. The key embeddings $M^k\in\mathbb{R}^{N\times{d_k}}$ are pre-trained to reflect the identity of KCs and their mutual relationships based on the question-KC bipartite graph. The value embeddings $M_t^{v}\in\mathbb{R}^{N\times{d_v}}$ represent the latest mastery state for each KC, and they are updated by observing the exercise answering sequence. Thus, our TGM forms a relational memory that captures the history of the exercise answering sequence and relationships between KCs as key similarity edges. Extracting information from the memory module is performed by two procedures including addressing and reading. We detail each of them as follows.

\textbf{Memory Addressing} to read knowledge mastery features relevant to a given question $q_t$, we perform associative addressing to the TGM by calculating relevancy read vector $w_t$ using the following equation:

\begin{equation}
    w_t(i)=\frac{exp({k^{q_t}}^{\intercal}k^{c^i})/\tau}{\sum_{j=1}^{N}exp({k^{q_t}}^{\intercal}k^{c^j})/\tau},\forall{i\in[1,N]}
    \label{eq:qkc_relevancy_vector}
\end{equation}

\noindent where $k^{q_t}$ is the key embedding of the question at time $t$, $k^{C^i}$ is the key embedding of KC $c^i$, $\tau\in[0,1]$ is a temperature parameter controlling the sharpness of the Softmax distribution, and $N$ is the total number of KCs.

\textbf{Memory Reading} we read relevant knowledge state aspects of $q_t$ from the current temporal graph memory $G_t$ using a key-value GCN module while attending to the relevancy read vector $w_t$. This module extends the aggregation calculations in conventional GCNs~\cite{GCNs17} by considering KC keys to calculate soft edges as self-attention weights between nodes. Instead of using a fixed adjacency as in the traditional GCN case, we use a dynamic adjacency matrix represented by the self-attention matrix~\cite{Vaswani17,shaw2018self}. We note that the KC keys matrix $M^k$ is initialized via the KC embedding pretraining stage and remains static. In contrast, the queries matrix $M_t^q\in\mathbb{R}^{N\times{d_q}}$ is dynamic and learned in an end-to-end manner with the answer prediction objective to compensate for the personal aspects of each student. The adjacency matrix at time point $t$ is calculated as follows:

\begin{equation}
    A_t=\Softmax(\frac{{M_t^q}{M^k}^\intercal}{\sqrt{d_k}})
    \label{eq:adj_calc}
\end{equation}

\noindent where $\sqrt{d_k}$ is the normalization factor quantified by the square root of dimensions of the key vector.

Afterward, we distill a read vector from the TGM using a variant of graph convolution operation~\cite{GCNs17} that attends node embeddings on the relevancy read vector $w_t$ before aggregation as follows:

\begin{equation}
    H^{j+1}=ReLU(\hat{D}^{-\frac{1}{2}}\hat{A}\hat{D}^{-\frac{1}{2}}\hat{H}^{j}W^{j})
    \label{eq:gcn_agg}
\end{equation}

\noindent where $ReLU(x)=max(0,x)$ is a non-linear activation function, $\hat{D}$ is the diagonal degree matrix calculated after performing binary masking on the adjacency matrix $A$ values by turning values less than the first quartile ($Q_1$) into $0$ and higher values into $1$, $\hat{A}=A+I$ is the adjacency matrix with self-loops where $I$ is the identity matrix, $\hat{H}^j=w_t\odot{H^j}$ is a scaled node feature matrix after performing row-wise multiplication with the question relevancy vector $w_t$, and $W^j$ is a learnable weight matrix for layer $j$ in the GCN. We note that $H^0=w_t\odot{M_t^{v}}$ is the first GCN layer input. We consider the cutoff threshold of the binary masking to calculate the degree matrix as a hyper-parameter quantified using empirical analysis.

We feed the output of the last GCN layer $H^l$ into a learnable non-linear layer to get the read vector $r_t\in\mathbb{R}^{d_v}$ summarizing $q_t$ relevant aspects of the current knowledge state.

\begin{equation}
    r_t=Tanh(W\cdot{flatten(H^l)}+b)
    \label{eq:tgm_read_vec}
\end{equation}

\noindent where $W$ and $b$ are the learnable weight matrix and bias vector of the $Tanh$ layer respectively .

\subsection{Exercise Sequence Context Estimation}
Our model's design assumes two forms of context to robustly estimate a question's mastery state, including long-term context (i.e., persists across time windows for the same student) represented by the TGM's read vector $r_t$ and a short-term context (i.e., resets across time windows for the same student) distilled from the exercise answering sequence in the current session represented by $h_t\in\mathbb{R}^{d_v}$.

To count for the exercise sequence context, we sequentially feed the mastery states of previous questions in the exercise sequence weighted by their relevancy to question $q_t$ to a gated recurrent unit (GRU) to get the sequence context $h_t$ for $q_t$. We first calculate the sequence relevancy vector $o_t$ as follows:

\begin{equation}
    o_t(i)=\frac{exp({k^{q_t}}^{\intercal}k^{q^i})/\tau}{\sum_{j=1}^{S}exp({k^{q_t}}^{\intercal}k^{q^j})/\tau},\forall{i\in[1,S]}
    \label{eq:qq_relevancy_vector}
\end{equation}

\noindent where $k^{q_t}$ is the key embedding of the question at time $t$, $k^{q^i}$ is the key embedding of question $q^i$, $\tau\in[0,1]$ is a temperature parameter controlling the sharpness of the Softmax distribution, and $S$ is the max time window length for questions to consider in the past exercise sequence. We report our empirical value for $S$ in Section~\ref{sec:implementation}.

Then, we calculate the sequence context $h_t$ as follows:

\begin{equation}
    h_t=GRU((o_t(i)U(i))_{i=1}^{S})
    \label{eq:seq_context}
\end{equation}

\noindent where $U=(u_t)_{t=1}^{S}$ is the sequence of past question mastery states encoded by their answer status (see Eq.~\ref{eq:tgm_update_vec}).


\subsection{Answer Prediction}
To predict the correct answer probability for question $q_t$, we concatenate long and short-term contexts to form the question mastery state $m_t=concat(r_t,h_t)$, then, we feed $m_t$ to a feed-forward answer prediction head to get the correct answer probability $\hat{a}_t\in[0,1]$. 

\begin{equation}
    \hat{a}_t=\sigma(W\cdot{m_t}+b)
    \label{eq:answer_prediction}
\end{equation}

\noindent where $\sigma$ is Sigmoid activation, $W$ is a learnable weight matrix, and $b$ is a learnable bias vector.

All of our model's learnable parameters are optimized by following back-propagation from a binary cross-entropy loss function:

\begin{equation}
    L=\frac{1}{Z}\sum_{i=1}^{Z}-(a_i\log{\hat{a}_i}+(1-a_i)\log{(1-\hat{a}_i)})
    \label{eq:loss}
\end{equation}

\noindent where $Z$ is a training mini-batch size, $a_i$ is the true answer, and $\hat{a}_i$ is the predicted answer.

\subsection{Memory Update}
After answer prediction, we update the TGM value embeddings $M_t^v$ representing the student's mastery level for each KC guided by the recent mastery state $m_t$, the status of answer prediction $(\hat{a}_t,a_t)$, and the current relevancy vector $w_t$. We represent the answer prediction status as one of four possible values based on the viable binary values for $(\hat{a}_t,a_t)$ including $\{(1,1),(1,0),(0,1),(0,0)\}$, with $1$ for correct answer and $0$ for incorrect answer. Accordingly, we use a static encoding matrix $A\in\mathbb{R}^{4\times{d_k}}$ to encode each answer status value and add the corresponding status vector to $m_t$ in a similar fashion of positional encoding in transformers~\cite{Vaswani17} to formulate the update vector $u_t$.
\begin{equation}
    u_t=A[(\hat{a}_t,a_t)]+m_t
    \label{eq:tgm_update_vec}
\end{equation}

We follow an associative memory update procedure~\cite{NTMs14} involving learnable erase and update gates; the former is responsible for erasing parts no longer relevant from memory, while the latter adds new important information. Only the value matrix ${M_t}^v$ of the TGM is updated after each time step to track changes in each KC's mastery state. Equations for the erase $e_t$ and add $z_t$ signals are as follows:

\begin{equation}
    e_t=\sigma(W_e\cdot{u_t}+b_e)
\end{equation}

\begin{equation}
    z_t=Tanh(W_z\cdot{u_t}+b_z)
\end{equation}

\noindent where $W_e$, $W_z$, $b_e$, and $b_z$ are learnable weight matrices and bias vectors for erase and add signals, respectively.

Given the erase and add signal vectors, we utilize them to update the KC value embedding matrix $M_t^v$ for the next time step $M_{t+1}^v$ as follows:

\begin{equation}
    M_{t+1}^{v}(i)=M_t^v(i)[1-w_t(i)e_t]+w_t(i)z_t
    \label{eq:tgm_state_update}
\end{equation}

To count for a student's forgetting behavior~\cite{forgetting_15}, we introduce a memory temporal decay mechanism that relies on question-KC relevancy in the exercise sequence inspired by the potential of memory decay approaches~\cite{NotAllMemSame21} to deal with long-term dynamics in sequences. Relevant forgetting-robust KT methods~\cite{abdelrahman2022deep,forget_NagataniZSCCO19} depend on hand-crafted features to represent a student's forgetting behavior. While effective, such representation might not generalize well from one scenario to another in addition to the need to model additional features besides ones imposed by the answer prediction objective. Alternatively, we utilize an adaptive memory exponential decay function with a decay ratio proportional to each KC's exercise sequence relevancy signal. We apply the memory decay function on all $M_{t+1}^v$ slots. Our memory decay function is defined as follows:

\begin{equation}
    \phi(M_{t+1}^v(i))=M_{t+1}^v(i)(1-\gamma)^{(1-w_t(i))}
    \label{eq:memory_decay}
\end{equation}

\noindent where $\phi(x)$ is the weight decay function, $\gamma\in(0,1]$ is the decay factor, and $(1-w_t(i))$ is the decay temporal step.

The design of the decay function in Equation~\ref{eq:memory_decay} poses decay steps that are proportional to each KC relevancy weight represented by $w_t$ thus, ones that were more relevant to exercises are decayed with a lower rate and vise versa. We use a decay value of $\gamma=0.02$ determined empirically on our experimental datasets.

\begin{algorithm}
\caption{TGMN Model Training}\label{alg:TGM}
\begin{algorithmic}[1]
\Initialize{
\begin{itemize}
    \item Question and KC embedding from TGMN pertaining procedure (Section \ref{sec:embedding_learning})
    \item Other weights, bias, and embedding parameters arbitrarily
\end{itemize}}\newline
\textbf{Begin:}
\For{epoch$\,\in\,1,\dots,n$}
  \For{mini-batch$\,\in\,1,\dots,m$}
  \ForAll{sample$\,(q_t,a_t)\,\in\,$mini-batch}
    \State{Embed input question $q_{t}$} 
    \State{Get relevance vector ${w_t}$ (Eq.~\ref{eq:qkc_relevancy_vector}) }
    \State{Get weighted TGM embedding ${H^l}$ (Eq.~\ref{eq:gcn_agg}) }
    \State{Get TGM read vector ${r_t}$ (Eq.~\ref{eq:tgm_read_vec})}
    \State{Get sequence context ${h_t}$ (Eq.~\ref{eq:seq_context})}
    \State{Get question mastery state ${m}_{t}=concat(r_t,h_t)$}
    \State{Predict answer $\hat{a}_t$ (Eq.~\ref{eq:answer_prediction})}
    \State{Get TGM update vector ${u_{t}}$ (Eq.~\ref{eq:tgm_update_vec})}
    \State{Update TGM state $M_{t+1}^{v}$ (Eq.~\ref{eq:tgm_state_update})}
    \State{Decay TGM state $\phi(M_{t+1}^v$ (Eq.\ref{eq:memory_decay})}
    \EndFor
    \State{Perform Adam~\cite{santoro16} (Eq.~\ref{eq:loss})}
  \EndFor
\EndFor
\end{algorithmic}
\end{algorithm}

\section{Experiments}
\label{sec:exp}
In this section, we present the experimental design to evaluate the effectiveness of our TGMN model, aiming to  address the following research questions: 
\begin{itemize}
\item[\textbf{RQ1}:] How does our proposed TGMN model compare to the state-of-the-art KT models on well-established KT benchmarks?
\item[\textbf{RQ2}:] What is the impact of different building blocks in our proposed model on its performance?
\item[\textbf{RQ3}:] To what extent does the pre-training procedure enhance embedding quality for questions and KCs? How does it compare to conventional embedding approaches in KT literature?
\item[\textbf{RQ4}:] What is the computational complexity of our model? How does it compare to other comparative KT models?

\end{itemize}

\subsection{Datasets}
\label{sec:datasets}

\begin{table}
  \caption{Statistics of the KT datasets utilized in our experimental design.}
  \label{tbl:datasets}
 \begin{adjustbox}{max width=\textwidth}
  \begin{tabular}{lcccc}
    \toprule
    \hspace*{0cm}Dataset&\#Students&\#Questions&\#Exercises&\#KCs \\
    \midrule
    ASSISTments2009&$4,151$&$110$&$325,637$&$110$\\
   Statics2011&$335$&$1,362$&$190,923$&$85$\\
    Synthetic-5&$4,000$&$50$&$200,000$&$5$\\
    Kddcup2010&$575$&$436$&$607,026$&$112$\\
    Eedi&$118,971$&$27,613$&$15,867,850$&$388$\\
    DBE-KT22&$1,361$&$212$&$167,222$&$98$\\
  \bottomrule
\end{tabular}
\end{adjustbox}
\end{table}

\begin{itemize}
\item\textbf{ASSISTments2009\footnote{ASSISTments2009:\url{  https://sites.google.com/site/assistmentsdata/home/assistment-2009-2010-data/skill-builder-data-2009-2010}}:} This dataset sourced from the ASSISTments online education platform in $2009-2010$, comprises $110$ unique school mathematics questions. It involved $4,151$ students, generating a dataset with $325,637$ question-answer pairs spanning $110$ KCs representing diverse mathematical concepts.

\item\textbf{Statics2011\footnote{Statics2011:\url{https://pslcdatashop.web.cmu.edu/DatasetInfo?datasetId=507}}:} This dataset sourced from a $2011$ engineering course at Carnegie Mellon University, features responses from $333$ students to a distinct set of $1,223$ questions. With $189,297$ exercises, including questions and corresponding answers, the dataset addresses $85$ KCs related to diverse engineering topics.

\item \textbf{Synthetic-5\footnote{Synthetic-5:\url{https://github.com/chrispiech/DeepKnowledgeTracing/tree/master/data/synthetic}}:} This dataset was created by the authors of DKT ~\cite{DKT2015}, simulates a learning experience with $4,000$ student agents responding to $50$ unique questions. This simulation results in $200,000$ exercises comprising question-answer pairs, and the dataset is structured around $5$ KCs representing underlying knowledge domains or topics within the learning material.

\item \textbf{Kddcup2010\footnote{Kddcup2010:\url{https://pslcdatashop.web.cmu.edu/KDDCup/downloads.jsp}}:} This dataset is derived from an algebra course conducted on the Cognitive Algebra Tutor system~\cite{kdd2010} during $2005-2006$, includes $436$ unique questions answered by $575$ students. With a total of $607,026$ exercises, encompassing both questions and their corresponding answers, the dataset encompasses $112$ KCs representing a variety of algebraic topics covered in the course.

\item\textbf{Eedi\footnote{Eedi:\url{https://eedi.com/projects/neurips-education-challenge}}:} 
This dataset released for the NeurIPS Education Challenge in 2021~\cite{neurips2020educationchallenge} is a comprehensive collection of mathematics question logs. The dataset contains $27,613$ unique questions and was completed by $118,971$ participating students, resulting in a collection of $15,867,850$ question and answer pairs and $388$ KCs. 

\item\textbf{DBE-KT22\footnote{DBE-KT22:\url{https://dataverse.ada.edu.au/dataset.xhtml?persistentId=doi:10.26193/6DZWOH}}:} This dataset consists of student exercise practicing in the Relational Databases course taught at the Australian National University (ANU) in Australia~\cite{abdelrahman2022dbe}. The dataset contains $212$ unique questions and was completed by $1,361$ participating students, resulting in a collection of $167,222$ question and answer pairs and $98$ KCs.

\end{itemize}

\subsection{Answer Prediction Performance Evaluation}

This experiment answers the research question \textbf{RQ1} by contrasting the performance of our TGMN model with the state-of-the-art KT models on six well-established datasets in the KT literature over five independent runs. The KT models for comparison cover both sequence-based and graph-based KT methods, including:

\begin{itemize}

\item \textbf{At-DKT}~\cite{atDKT23}: This model extends the DKT~\cite{DKT2015} model with two auxiliary tasks to predict the related KCs for a given question and its past mastery state. 

\item  \textbf{SKVMN}~\cite{SKVMN}: This model utilizes a key-value memory and Hop-LSTM to capture sequential dependencies among questions in a sequence of interactions. This approach allows the model to update students' knowledge based on their responses to relevant questions.
 
\item \textbf{SAKT}~\cite{SAKT}: This model follows an attention mechanism~\cite{BahdanauCB14} to focus on the most relevant parts of the student's sequence of interactions and give them higher weight during the prediction process.

\item \textbf{AKT}~\cite{AKT20}: This model combines an attention model with Rasch model-based embeddings. The attention mechanism assigns weights to the questions in a sequence, indicating their importance for predicting the current question. Additionally, the attention weights are exponentially decayed based on the distance between the questions in the sequence.

\item \textbf{GKT}~\cite{GKT19}: This model incorporates a graph neural network (GNN) to extract information that captures dependencies across questions. A graph neural network is a neural network specifically designed to operate on graph-structured data.

\item\textbf{SKT}~\cite{TongLHHCLM020}: This model aims to capture multiple KCs relations, including similarity and prerequisite relations. Incorporating these relations into the model seeks to capture the interdependencies and connections between different KCs.

\item  \textbf{GIKT}~\cite{GIKT20}: This model utilizes the relationship between questions and KCs, represented as a graph structure, to learn effective embeddings for answer prediction and to capture the connections and interactions between questions and the corresponding KCs.

\item\textbf{DGMN}~\cite{abdelrahman2022deep}: This model creates a dynamic graph to represent relationships between KCs and represents forgetting behavior throughout a KC, which has the benefit of capturing inverse correlations between questions.

\item  \textbf{PEBG}~\cite{PEBG}: This model emphasizes the acquisition of pre-trained exercise embeddings to improve the accuracy of knowledge tracing (KT) tasks. It achieves this by representing exercises, exercise difficulties, and KCs in a unified manner using a bipartite graph.

\end{itemize}

We utilize the optimal configuration recommended by the original work for the baseline techniques' hyper-parameters (e.g., input time window length) to evaluate the dataset. If the dataset was not included in the initial work, we conduct hyper-parameter tuning using a randomly sampled held-out set of $5\%$ of the original dataset size and select the optimal configuration for comparison. We perform a 5-fold splitting strategy for train-test splits and report the averaged AUC results. Also, we perform the statistical significance \emph{t-test} on the results with \emph{p-value}$\mathrm{=0.05}$.

\begin{figure*}[t!]
\includegraphics[width=1\textwidth, height=4.5cm]{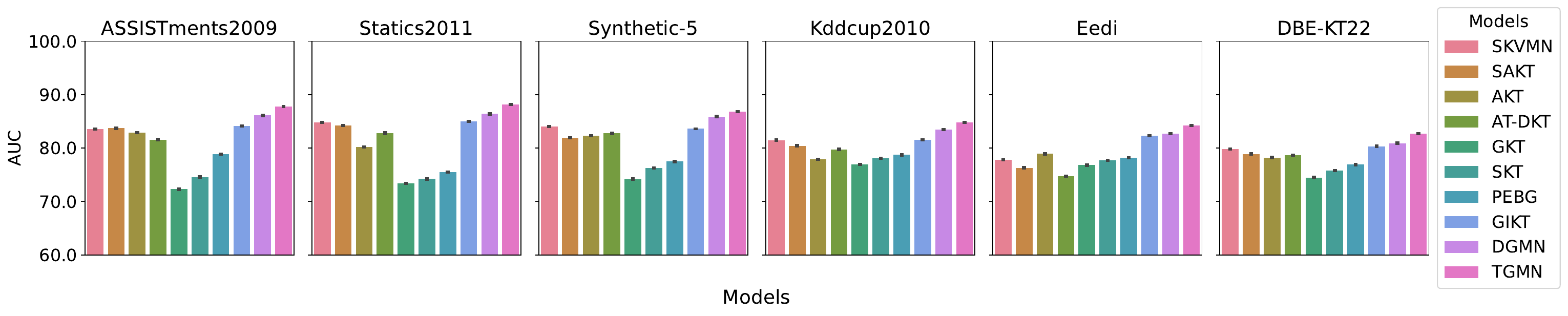}
\caption{The Area under ROC curve (AUC) metric results comparing TGMN with the state-of-the-art KT models over all the datasets. We depict the average with the standard deviation of the metric values using a 5-fold splitting strategy (best in colors).}
\label{fig:AUC_res}\vspace{-0.1cm}
\end{figure*}

Table~\ref{tbl:AUC} summarizes the average accuracy results, while Figure~\ref{fig:AUC_res} visualizes average AUC results using bar plots. We observe that TGMN significantly outperforms all the other models on both metrics across all the datasets. TGMN achieves an accuracy improvement over the nearest top-performing model (DGMN) by a margin of $\mathrm{1.7\%}$, $\mathrm{1.8\%}$, $\mathrm{0.9\%}$, $\mathrm{1.4\%}$, $\mathrm{1.5\%}$, and $\mathrm{1.8\%}$ on the datasets \emph{ASSISTments2009}, \emph{Statics2011}, \emph{Synthetic-5}, \emph{Kddcup2010}, \emph{Eedi}, and \emph{DBE-KT22}, respectively. These results are statistically significant (\emph{p-value}$\mathrm{<0.05}$) based on the \emph{t-test} conducted over five independent (i.e., different seeds) runs. We also note that sequence-based KT models (first section of Table~\ref{tbl:AUC}) are generally performing better than graph-only methods (second section of Table~\ref{tbl:AUC}), yet hybrid models (third section of Table~\ref{tbl:AUC}) using both sequence and graph information are leading the performance results with significant margins from sequence-only or graph-only models. This finding confirms the effectiveness of fusing features from both exercise answering sequence and KC-question relationship graphs in enhancing the answer prediction.

 \begin{table*}
  \caption{Average accuracy results over five independent runs comparing our model with the state-of-the-art KT models over all the datasets.}
  \centering
  \label{tbl:AUC}
  \begin{tabular}{l|cccccc}
    \toprule
     \multirow{2}{*}{Model}&\multicolumn{6}{c}{Dataset}\\\cline{2-7}
   &ASSISTments2009&Statics2011&Synthetic-5&Kddcup2010&Eedi&DBE-KT22\\
      \midrule
      
      SKVMN~\cite{SKVMN}&$83.6\pm0.012$&$84.8\pm0.009$&$84.0\pm0.022$&$81.5\pm0.008$&$77.8\pm0.004$&$79.8\pm0.063$\\
      SAKT~\cite{SAKT}&$83.7\pm0.003$&$84.2\pm0.015$&$81.9\pm0.031$&$80.4\pm0.003$&$76.3\pm0.005$&$78.9\pm0.041$\\
      AKT~\cite{AKT20}&$82.9\pm0.018$&$80.2\pm0.017$&$82.3\pm0.038$&$77.9\pm0.007$&$78.9\pm0.008$&$78.2\pm0.051$\\
      AT-DKT~\cite{atDKT23}&$81.6\pm0.026$&$82.8\pm0.004$&$82.7\pm0.017$&$80.8\pm0.005$&$74.7\pm0.011$&$77.6\pm0.031$\\\hline
      GKT~\cite{GKT19}&$72.3\pm0.001$&$73.4\pm0.011$&$74.2\pm0.027$&$76.9\pm0.002$&$76.8\pm0.081$&$74.5\pm0.002$\\
      SKT~\cite{TongLHHCLM020}&$74.6\pm0.003$&$74.2\pm0.001$&$76.3\pm0.014$&$78.1\pm0.006$&$77.7\pm0.022$&$75.8\pm0.031$\\
      PEBG~\cite{PEBG}&$78.9\pm0.001$&$75.5\pm0.033$&$77.5\pm0.006$&$78.7\pm0.028$&$78.2\pm0.039$&$76.9\pm0.011$\\\hline
      GIKT~\cite{GIKT20}&$84.1\pm0.009$&$85.0\pm0.003$&$83.6\pm0.018$&$81.5\pm0.012$&$82.3\pm0.002$&$80.3\pm0.005$\\
      DGMN~\cite{abdelrahman2022deep}&$86.1\pm0.001$&$86.4\pm0.013$&$85.9\pm0.006$&$83.4\pm0.012$&$82.7\pm0.005$&$80.9\pm0.003$\\
      TGMN (ours)&$\mathbf{87.8\pm0.003}$&$\mathbf{88.2\pm0.015}$&$\mathbf{86.8\pm0.002}$&$\mathbf{84.8\pm0.011}$&$\mathbf{84.2\pm0.007}$&$\mathbf{82.7\pm0.016}$\\
    \bottomrule
\end{tabular}
\end{table*}

\subsection{Ablation Study}

We conduct an ablation study to answer the research question \textbf{RQ2}. We compare different variants of our TGMN model to evaluate the impact of different building blocks on its performance. These variants are as follows:
\begin{itemize}
    \item \textbf{Base}: This variant only includes a basic Temporal Graph Memory without using Sequence Context, Memory Decay, and Answer Encoding modules.

    \item \textbf{TGM-SC}: This variant includes a basic Temporal Graph Memory and Sequence Context without using  Memory Decay and Answer Encoding modules.
    
    \item \textbf{TGM-F}: This variant includes a basic Temporal Graph Memory, Sequence Context, and Memory Decay without using the Answer Encoding module.
    
    \item \textbf{TGMN}: This variant includes all the components of TGMN.
    
\end{itemize}

Table~\ref{tbl:ablation} summarizes the average accuracy results for different variants of the TGMN model. Based on Table~\ref{tbl:ablation}, we have the following findings: 

\begin{itemize}

    \item[(1)] The impact of the Sequence Context module can be observed by comparing Base and TGM-SC. A statistically significant (\emph{p-value}$\mathrm{<0.05}$) performance enhancement by a margin of $\mathrm{1.8\%}$, $\mathrm{1.5\%}$, $\mathrm{1.2\%}$, $\mathrm{1.4\%}$, $\mathrm{1.3\%}$ and $\mathrm{1.2\%}$ is achieved for \emph{ASSISTments2009}, \emph{Statics2011}, \emph{Synthetic-5}, \emph{Kddcup2010}, \emph{Eedi}, and \emph{DBE-KT22}, respectively.
    
    \item[(2)] The impact of the Memory Decay module can be observed by comparing TGM-SC with TGM-F. A significant performance margin of $\mathrm{2.0\%}$, $\mathrm{1.4\%}$, $\mathrm{2.4\%}$, $\mathrm{2.0\%}$, $\mathrm{0.9\%}$, and $\mathrm{0.9\%}$ exists between these two models for \emph{ASSISTments2009}, \emph{Statics2011}, \emph{Synthetic-5}, \emph{Kddcup2010}, \emph{Eedi}, and \emph{DBE-KT22}, respectively.

    \item[(3)] The impact of using an Answer Encoding on the performance can be seen by comparing TGM-F against TGMN. Significant performance margins of $\mathrm{2.0\%}$, $\mathrm{2.1\%}$, $\mathrm{1.3\%}$,  $\mathrm{1.8\%}$, $\mathrm{1.4\%}$, and $\mathrm{1.8\%}$ are obtained by DGMN-Basic for datasets \emph{ASSISTments2009}, \emph{Statics2011}, \emph{Synthetic-5}, \emph{Kddcup2010}, \emph{Eedi}, and \emph{DBE-KT22}, respectively

\end{itemize} 

\begin{table*}
  \caption{Average accuracy results over five independent runs comparing different variants of the TGMN model in our ablation study.}
  \label{tbl:ablation}
\centering
  \begin{tabular}{ll|ccc|c}
  \hline
   Model&& Base &  TGMN-SC & TGMN-F & TGMN \\\hline
 \multirow{4}{*}{Component} & TGM & \checkmark&\checkmark&\checkmark&\checkmark\\ 
 &Sequence Context&$\times$& \checkmark&\checkmark&\checkmark\\
&Memory Decay&$\times$&$\times$&\checkmark&\checkmark\\
&Answer Encoding&$\times$&$\times$&$\times$&\checkmark\\\hline
 \multirow{6}{*}{Dataset}&ASSISTments2009&$82.0\pm0.001$&$83.8\pm0.011$&$85.8\pm0.003$&$\mathbf{87.8\pm0.003}$\\
&Statics2011&$83.2\pm0.004$&$84.7\pm0.023$&$86.1\pm0.035$&$\mathbf{88.2\pm0.015}$\\
&Synthetic-5&$81.9\pm0.018$&$83.1\pm0.052$&$85.5\pm0.003$&$\mathbf{86.8\pm0.002}$\\
&Kddcup2010&$79.6\pm0.002$&$81.0\pm0.022$&$83.0\pm0.007$&$\mathbf{84.8\pm0.011}$\\
&Eedi&$80.6\pm0.004$&$81.9\pm0.016$&$82.8\pm0.003$&$\mathbf{84.2\pm0.007}$\\
&DBE-KT22&$78.8\pm0.006$&$80.0\pm0.013$&$80.9\pm0.008$&$\mathbf{82.7\pm0.016}$\\\hline
\end{tabular}
\end{table*}  
     
\subsection{Embedding Quality Analysis}     
In this experiment, we answer \textbf{RQ3} by performing a training time performance comparison across four TGMN variants, each using a different embedding initialization. This design assumes that effective embedding initialization will result in better training performance and faster convergence than inferior ones. In this experiment, we utilize the \emph{DBE-KT22}~\cite{abdelrahman2022dbe} dataset for question and KC text information availability. The four variants are as follows:
\begin{itemize}
    \item \textbf{Text+PT}: this variant initializes question and KC embedding using the BERT~\cite{Vaswani17} language model embeddings from their text tags and performs our proposed pre-training procedure introduced in Section~\ref{sec:embedding_learning} that further enhances the embedding space by leveraging mutual relationships in the question-kc graph of the dataset.
    \item \textbf{OHC+PT}: this variant initializes question and KC embedding using one-hot encodings and follows our proposed pre-training procedure. 
    \item \textbf{Text}: this variant only uses the BERT~\cite{Vaswani17} language model embeddings from questions and Kcs text tags to initialize their embedding representation.
    \item \textbf{OHC}: this variant only uses one-hot encodings to initialize question and KCs embedding representation.
\end{itemize}

\begin{figure}[htbp]
\centerline{\includegraphics[scale=0.4]{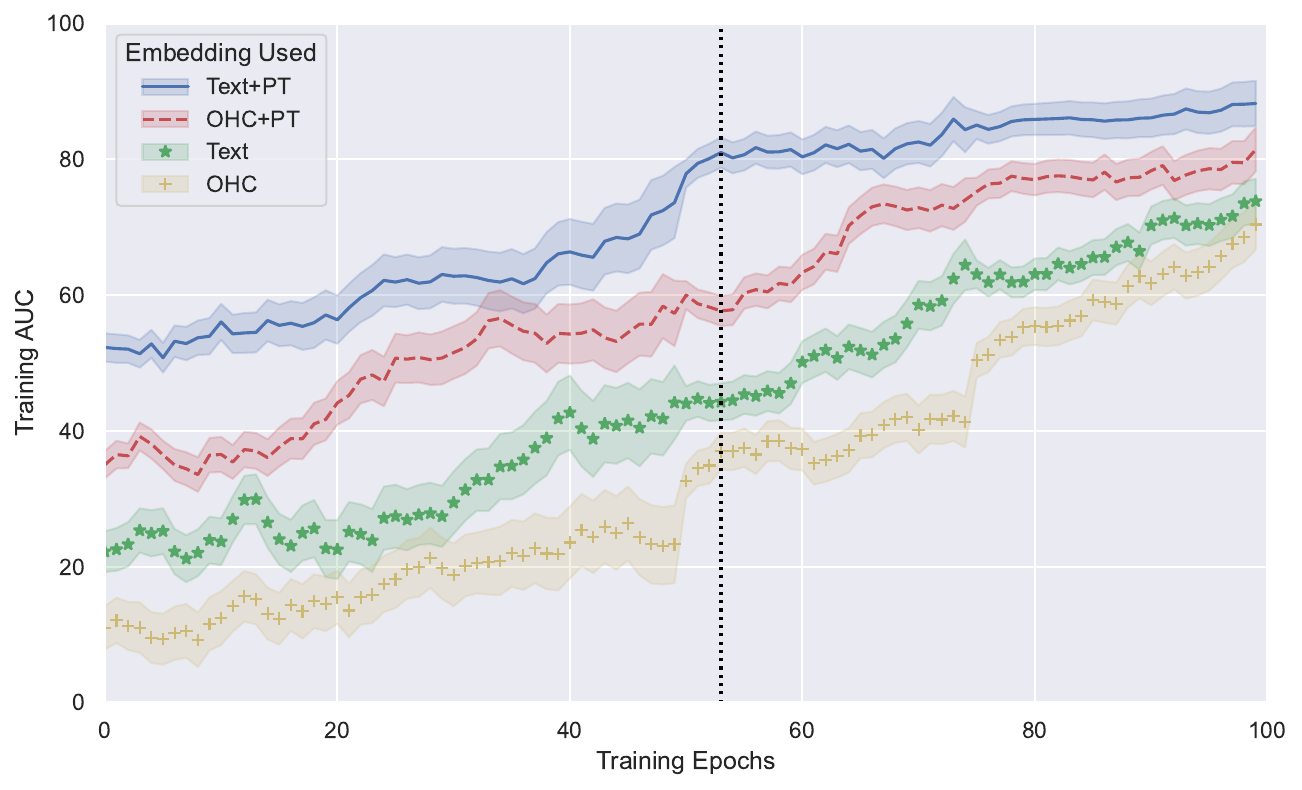}}
\caption{The Area under ROC curve (AUC) training curves comparing four TGMN variants, each using a different embedding initialization. }
\label{fig:embedding_analysis}
\end{figure}

\begin{figure*}[htbp]
\centerline{\includegraphics[scale=0.5]{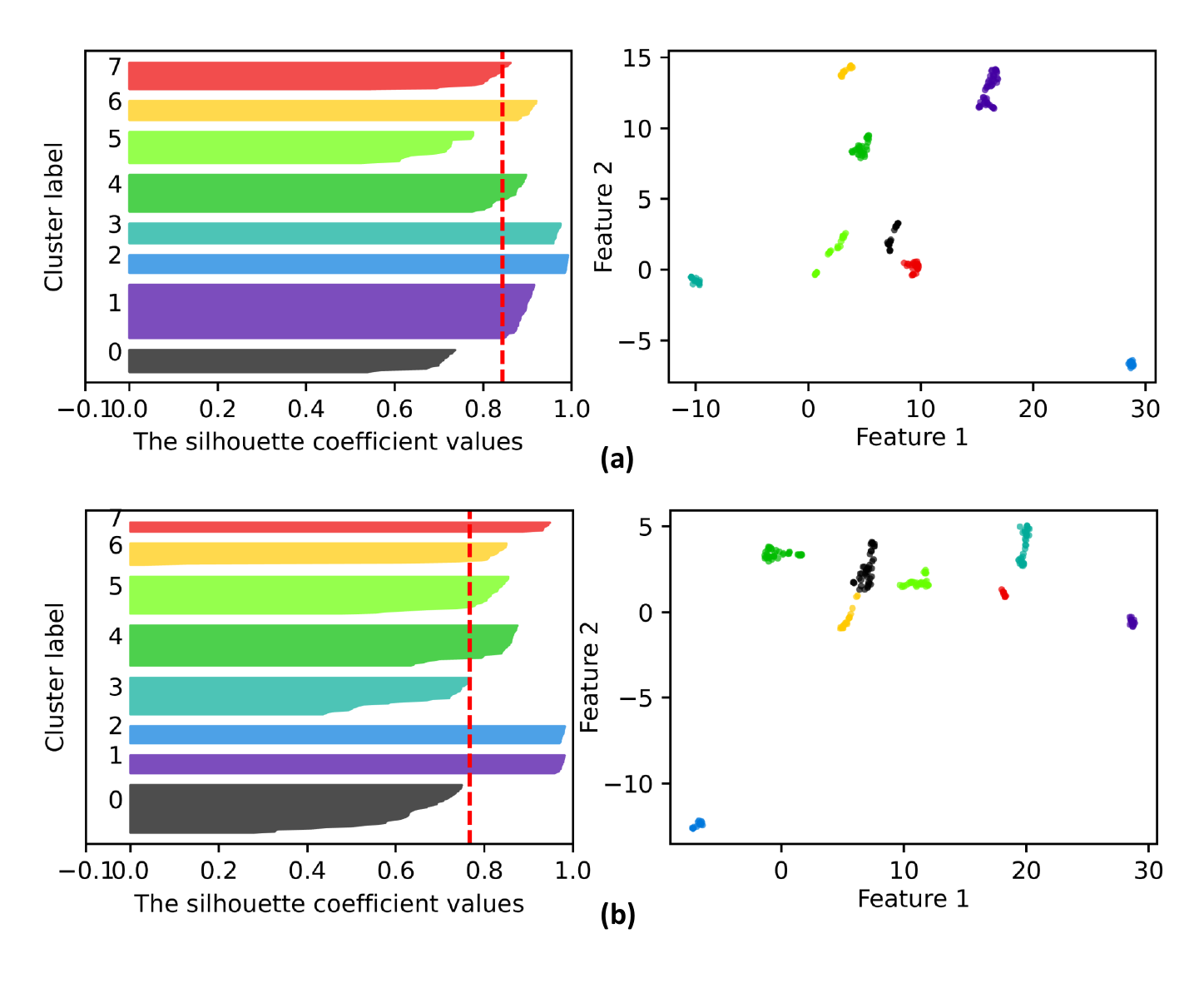}}
\caption{Clustering evaluation results on the DBE-KT22~\cite{abdelrahman2022dbe} dataset showing the silhouette coefficient values on the left side and t-SNE~\cite{van2008visualizing} embeddings on the right side. (a) The \emph{Text+PT} variant. (b) The \emph{Text}-only variant.}
\label{fig:clustering_analysis}
\end{figure*}

Figure~\ref{fig:embedding_analysis} depicts TGMN training AUC curves using different embedding methods over $100$ training epochs. We observe that the \emph{Text+PT} outperformed others in terms of the final AUC value of $90$\% and it converged faster (dotted line at epoch 53) with a large margin from the nearest variant (i.e., \emph{OHC+PT}) of $20$ AUC points. By contrasting the curves of the \emph{Text+PT} and \emph{OHC+PT} variants, we can observe the impact of using text embedding compared to one-hot encodings usually used in KT literature. The effectiveness of our proposed embedding pre-training procedure (see Section~\ref{sec:embedding_learning}) could be highlighted by comparing curves of variants ending with \emph{PT} suffix to the \emph{Text}-only and \emph{OHC}-only ones. An interesting finding is that the variant starting with one-hot encodings and then following our pre-training could still outperform and converge faster than the one using text embeddings without following our pre-training objective (i.e., \emph{Text}). Figure~\ref{fig:clustering_analysis} shows K-means~\cite{kmean_2003} clustering results comparing the \emph{Text+PT} (silhouette coeff=$0.85$) and the \emph{Text}-only (silhouette coeff=$0.76$) variants on the DBE-KT22 dataset~\cite{abdelrahman2022dbe}. The better silhouette coefficient achieved by the \emph{Text+PT} variant highlights the impact of our pre-training method. The number of clusters was empirically selected using the elbow method~\cite{elbow_method_2018}, and we map embeddings on 2D feature space using t-SNE~\cite{van2008visualizing}.

\subsection{Complexity Analysis}    

To answer the research question \textbf{RQ4}, we asymptotically calculate the time and space computational complexity for sample sequence-based models, graph-based models, hybrid models, and our proposed TGMN model. Table~\ref{tbl:comana} compares training computational and memory complexity for a subset of KT models with the TGMN model. $|Q|$ refers to the total number of questions, $v$ is the input embedding dimension, $S$ is the input sequence length, $N$ is the total number of KCs, $h$ is the recurrent memory's hidden state dimensions, $H$ is the total number of attention heads used, $a_k$ is the attention key dimension, and $k$ is the attention key dimension. While both of the \emph{DGMN} and \emph{TGMN} models utilize sequence and graph information, our model's time complexity is better as we follow answer encoding (i.e., vector summation) instead of doubling the question encodings $2|Q|$ to represent possible answer boolean states (i.e., true or false).

\begin{table}
  \caption{Training time and memory complexity comparison for a sample sequence-based, graph-based, and hybrid (including ours) KT models.}
  \label{tbl:comana}
 \resizebox{\columnwidth}{!}{
  \begin{tabular}{lcc}
    \toprule
    \hspace*{0cm}Method&Time Complexity& Memory Complexity \\
    \midrule
    SKVMN&$\mathcal{O}(2|Q| \times v)$&$\mathcal{O}(2N \times v)$\\
    SAKT&$\mathcal{O}(S^2 \times k)$&$\mathcal{O}(H \times a_k)$\\
    AKT&$\mathcal{O}(2S^2 \times k)$&$\mathcal{O}(2H \times a_k)$\\\hline
    GKT&$\mathcal{O}(|Q|^2 \times v)$&$\mathcal{O}(N^2)$\\
    DGMN&$\mathcal{O}(2|Q| \times v)$&$\mathcal{O}(N^2)$\\\hline
    TGMN (ours)&$\mathcal{O}(|Q| \times v)$&$\mathcal{O}(N^2)$ \\
  \bottomrule
\end{tabular}
}
\end{table} 

\section{Implementation Details}
\label{sec:implementation}
In this section, we present the details of implementing our TGMN model. For the embedding learning model, we use a three-layer feed-forward neural network with a \emph{ReLU} activation for the hidden layers and a liner activation at the final layer for the number of hops prediction head. For the backbone, we either use a frozen BERT model if text data is present in the dataset or add an embedding layer in case of one-hot encoding. For TGM module, we set $d_k$, $d_v$, and $d_q$ dimensions to be $512$. The question and KC key embeddings are initialized with the last hidden layer output of the embedding learning model. While the value $M^v$ and query $M^q$ matrices for KCs are initialized using a zero-mean random Gaussian distribution $N(0,\sigma)$. We use two GCN layers for the key-value GCN module, extending the implementation from the \emph{Pytorch Geometric} library~\footnote{https://pytorch-geometric.readthedocs.io/}. We use a two-layer GRU cell with a hidden state size of $512$ and recurrent dropout of $0.2$ for the sequence context module. We empirically (using held-out sets) set our time window length to be $15$ and the value of $0.02$ for the decay factor $\gamma$ of our memory decay function. Our model's learnable parameters are optimized end-to-end following the \emph{Adam} gradient decent algorithm \cite{santoro16}. We utilize one workstation machine with two Nvidia RTX Titian GPUs, each with $24$GB of dedicated memory, an Intel Core i9-13900k CPU, and a RAM size of $128$GB.

\section{Related Work}
\label{sec:rl}
The Knowledge Tracing (KT) literature could be generally categorized based on the type of context used for answer prediction. We identify three different context types: temporal context from the past exercise answering sequence, structural context from graphs representing relationships between questions and KCs, and hybrid contexts combining temporal and structural data via different fusion techniques. We explore each of those categories as follows.

\subsection{KT Methods Using Temporal Context} 
One of the early attempts to use deep neural networks for modeling the temporal context in the KT problem was the DKT~\cite{DKT2015} model. At its core, DKT uses a recurrent neural network (RNN) to memorize important information from the past exercise answering sequence representing the knowledge state for answer prediction. Inspired by attention methods~\cite{BahdanauCB14,Vaswani17}, the DKVMN~\cite{DKVMN17} model uses a key-value memory to store embeddings and knowledge state dynamics for multiple KCs. Answer prediction is achieved by attending question embeddings on the key-value memory content, which significantly enhances the prediction performance compared to the vanilla RNN used by the DKT model~\cite{DKT2015}. The Sequential Key-Value Memory Networks (SKVMN)~\cite{SKVMN} extended on the DKVMN model by adding a hop-LSTM layer above the key-value memory to aggregate the temporal context of related questions in the exercise answering sequence during answering prediction. Building on the \emph{Transformer} architecture~\cite{Vaswani17}, various studies~\cite{SAKT,AKT20,SAINT+21} have explored integrating attention mechanisms into KT models. Despite differences in their implementations, these studies share a core objective: to determine the attention weights of questions within a sequence of interactions, thereby reflecting their relative importance in predicting the likelihood of correctly answering the next question. This approach addresses a limitation of DKT, which assumes all questions in a sequence are equally significant.

\subsection{KT Methods Using Structural Context}
The structural context in KT tasks typically includes various relationships, such as the dependencies between knowledge components (KCs) and the connections between questions and their associated KCs. Graph-based KT methods aim to exploit this context to enhance answer prediction. The Graph Knowledge Tracing (GKT)~\cite{GKT19} model used a GNN~\cite{GNN08} network to aggregate the structural context from the question-KC graph for answer prediction. The Structure-based Knowledge Tracing (SKT)~\cite{TongLHHCLM020} model extended on GKT~\cite{GKT19} by assuming a dynamic graph structure where question-KC relationships could change over time and used Gated Recurrent Unit (GRUs)~\cite{GRU14} memory to store temporal structural changes. The Pre-training Embeddings via Bipartite Graph (PEBG)~\cite{PEBG} model focused on learning effective question embeddings by designing pre-training objectives on side information like question difficulty and relationships between questions and KCs represented by a bipartite graph. More specifically, the PEBG method follows contrastive learning~\cite{ChopraHL05} objectives to identify question similarities and KC similarities along with question-KC edge prediction. The authors showed that using such pre-training could significantly enhance existing models such as DKT~\cite{DKT2015} and DKVMN~\cite{DKVMN17}.

\subsection{KT Methods Using Hybrid Context}
Utilizing both temporal and structural contexts in the KT problem facilitates mitigating information gaps in each of them and enhances the overall performance. Exploiting this fact, multiple KT methods aimed at proposing novel fusion approaches for combining structural and temporal information in the KT domain. The Graph-based Interaction Knowledge Tracing (GIKT)~\cite{GIKT20} model combined multiple ideas, such as using RNNs as in the DKT~\cite{DKT2015} model, aggregating information from relevant questions in the past exercise sequence as in the SKVMN~\cite{SKVMN} model, and using graph neural networks as in the GKT~\cite{GKT19} model. GIKT inference workflow goes as follows: firstly, a graph convolution network (GCN)~\cite{GCNs17} is used to aggregate question embeddings from the question-KC graph to form the structural context; then, related question embeddings in the answering sequence are combined using an RNN to formulate a combined temporal and structural context. Comparison against temporal-only and structural-only KT models showed a significant performance edge for the GIKT model. The Deep Graph Memory Network (DGMN)~\cite{abdelrahman2022deep} is considered to be the state-of-the-art in the hybrid context approaches as it not only provides a novel way of fusing structural and temporal contexts but also counts for student forgetting behavior~\cite{forgetting_15} during the answer prediction. Similar to key-value memory approaches~\cite{DKVMN17,SKVMN}, the DGMN utilizes a key-value memory for capturing the temporal context from the past answering sequence, but it employs this memory to automatically build a KC-KC graph capturing structural context between involved KCs. The temporal context is fused with the structural one via a forget gate that explicitly models the student's forgetting features in terms of KC practice frequencies. One common drawback of these hybrid models is keeping separate representations for temporal and structural contexts and fusing them at a later stage. In contrast, our DGMN model utilizes a unified representation that captures joint embedding for structural and temporal information, which reduces its memory fingerprint and enhances the fusion quality.

\section{Conclusion}
\label{sec:conclusion}
In this work, we introduced a novel knowledge tracing model named temporal graph memory networks (TGMNs), capable of learning a unified representation for exercise answering context and KC-KC structural relationships context for an accurate answer prediction. Moreover, we count for the impact of a student's forgetting behavior during the answer prediction via a generic temporal memory decay technique without dependency on handcrafted forgetting features. Experimental results comparing the state-of-the-art KT models across multiple datasets showed a significant performance edge for the TGMN model. In addition, we performed an ablation study to assess the impact of each building block in our proposed model. Future work directions include exploring other ways to effectively distill information from the temporal graph memory, such as relational transformer architectures~\cite{DiaoL23} and learning with both supervised (e.g., answer prediction) and self-supervised objectives (e.g., mutual information maximization based on structural priors) during the answer prediction to enhance the knowledge tracing performance. 

\bibliographystyle{IEEEtran}
\bibliography{references}


\vfill

\end{document}